\documentclass[prl,twocolumn,showpacs,preprintnumbers,amsmath,amssymb]{revtex4}
\usepackage{graphicx}
\usepackage{dcolumn}
\usepackage{bm}

\begin{document}

\input{epsf}
\preprint{mpi-pks/0211018}

\title{Phenomenological theory of spinor Bose-Einstein condensates}

\author{Qiang Gu}

\affiliation{Max-Planck-Institut f\"{u}r Physik komplexer Systeme, 
N\"{o}thnitzer Stra{\ss}e 38, 01187 Dresden, Germany}

\date{\today}

\begin{abstract}
It was reported that Bose-Einstein condensation induces a spontaneous 
magnetization in spinor bosons. This phenomenon is called Bose-Einstein 
ferromagnetism (BEF). We propose a phenomenological model consistent with 
the prediction of BEF and show that BEF might be attributed to the intrinsic 
magnetic moment of particles. Taking BEF into account, the phase diagram of 
an optically trapped spinor condensate is reexamined.  
\end{abstract}


\maketitle

Alkali-metal atoms, such as $^{87}$Rb, $^{23}$Na, $^7$Li, in which 
Bose-Einstein condensation (BEC) has been experimentally 
realized\cite{cornell}, have an internal degree of freedom attributed to the 
hyperfine spin $F$. Accordingly, the order parameter of Bose condensed alkali 
atoms could have $2F+1$ components. Thus, although the superfluid 
properties of this system are considered to be essentially the same as those 
of superfluid $^4$He, one can expect that the hyperfine spin degrees of 
freedom could bring about remarkable differences between the two systems. 

Earlier experiments leading to BEC in alkali atoms were performed in magnetic 
traps. A relatively strong external magnetic field $H$ was applied 
to confine the BEC system. Because the atomic spin direction adiabatically 
followed $H$, the spin degree of freedom was frozen. As a result, the alkali 
atoms behaved like scalar particles although they carried spin\cite{leggett}. 
In 1998, Stamper-Kurn {\sl et al.} succeeded in cooling $^{23}$Na in a purely 
optical trap and achieved BEC\cite{ketterle}. With $H$ sufficiently small, 
the spin degree of freedom can become active and the spinor nature of the 
condensate can be manifested\cite{kett2}.  

The optical trap provides great opportunities to investigate spinor Bose gases 
experimentally. Furthermore, it stimulates enormous theoretical interest in 
studying various spin-related properties, e.g. 
the magnetism\cite{ho,ohmi,pu}, of the gases. For an $F=1$ Bose gas, the 
effective interactions $V({\bf r})$ between the atoms were derived 
as\cite{ho, ohmi}, 
\begin{eqnarray}
V({\bf r})= \frac {c_0}2 \psi^\dag_\sigma \psi^\dag_{\sigma'} \psi_{\sigma'} 
  \psi_\sigma + \frac {c_2}2 \psi^\dag_\sigma 
  \psi^\dag_{\sigma'}{\bf F}_{\sigma\gamma}
  \cdot {\bf F}_{\sigma'\gamma'} \psi_{\gamma'} \psi_{\gamma} ~,
\end{eqnarray}
where $\psi_{\sigma}({\bf r})$ is the field annihilation operator for an atom 
in hyperfine state $|F,\sigma\rangle$ ($\sigma=1,0,-1$) at position ${\bf r}$. 
Repeated sub-indices represent summation taken over all the hyperfine states. 
Besides a spin-independent scattering, a Heisenberg-like exchange interaction 
between hyperfine spins appears in the Hamiltonian, where the components of 
${\bf F}=(F_x, F_y, F_z)$ are $3\times 3$ spin matrices. According to the 
pioneering work of Ho\cite{ho} and Ohmi and Machida\cite{ohmi}, this term 
determines the phase diagram of the spinor condensate: the ground state can be 
either ferromagnetic or ``polar", depending on whether $c_2$ is negative or 
positive, respectively. However, recently Eisenberg and Lieb (referred to as EL 
hereafter)\cite{eisenberg} claimed that this scenario ignored an intrinsic 
{\sl ferromagnetism} of spinor bosons.

The intrinsic ferromagnetism in spin-$1$ ideal bosons was predicted by 
Yamada\cite{yamada} and recently studied in the context of atomic Bose-Einstein 
condensation by Simkin and Cohen\cite{Simkin}. They calculated the 
magnetization $M(H)$ and found that once BEC takes place, $M$ remains finite 
even if $H=0$, suggesting the system is magnetized spontaneously. This 
phenomenon is called Bose-Einstein ferromagnetism (BEF). 
Caramico D'Auria {\sl et al.} pointed out that the coexistence of BEC with BEF 
also appears in {\sl interacting} bosons\cite{caramico}. Moreover, rigorous 
proofs presented by S\"{u}t\H{o}\cite{suto} and EL\cite{eisenberg} show 
that a polarized state is among the degenerate ground state of spinor bosons. 
EL studied a Bose model with a {\sl spin-independent}, totally symmetric 
interaction $v$ between particles,
\begin{eqnarray}
H=-\sum_{i=1}^{N} \frac {\hbar^2}{2m} \nabla_i^2 + v({\bf r}_1,{\bf r}_2,...,
{\bf r}_N),
\end{eqnarray}
where ${\bf r}_i$ denotes the spatial coordinate of particle $i$. The specific 
model of Caramico D'Auria {\sl et al.} has the same symmetry as the above 
Hamiltonian.

In this communication, we propose a phenomenological model for the spinor Bose 
condensate. First of all, this model shows in a simple way that BEF can really 
take place at the same critical temperature as BEC. The results obtained are 
consistent with available theories. Then the phase diagram of optically trapped 
alkali atoms is 
reexamined, taking BEF into account. Furthermore, we try to understand the 
nature of BEF by analyzing the phenomenological model. 

To begin with, we derive an appropriate Ginzburg-Landau free energy density 
functional for second-order phase transitions that describes the coexistence of 
BEC and BEF. Generally, such a free energy density consists of three 
different parts: $f_t= f_b + f_m + f_c$, corresponding to the Bose condensed 
phase, the ferromagnetic phase and the coupling between the two phases (we omit 
the free energy of the normal Bose gas). Following Ginzburg and 
Pitaevskii\cite{ginzburg}, the free energy density of an {\sl isotropic} 
spin-$F$ Bose-Einstein condensate is given as
\begin{eqnarray}
f_b = \frac {\hbar^2}{2m} \nabla {\bf \Psi}^{\dag} \cdot \nabla {\bf \Psi}
    - \alpha |{\bf \Psi}|^2 + \frac {\beta}2 |{\bf \Psi}|^4  \nonumber\\ 
    + \frac {\beta_{s}}2 
      \Psi^\dag_\sigma \Psi^\dag_{\sigma'}{\bf F}_{\sigma\gamma} \cdot 
      {\bf F}_{\sigma'\gamma'} \Psi_{\gamma'} \Psi_{\gamma} ~,
\end{eqnarray}
where ${\bf \Psi}^{\dag}\equiv ({\bf \Psi}^T)^{*}=(\Psi_F^{*}, \Psi_{F-1}^{*},
..., \Psi_{-F}^{*})$ is the complex order parameter 
and $|{\bf \Psi}|^2={\bf \Psi}^{\dag}{\bf \Psi}$. Here the order parameter is 
defined as $\Psi_\sigma=\langle \psi_\sigma \rangle$, but we note that other 
forms of the order parameter are possible (see Leggett\cite{leggett}).  
The free energy density should have the same symmetry as the underlying 
Hamiltonian. The first three terms describe a system without any spin-dependent 
interactions, as given by Eq. (2). The fourth term has SU(2) symmetry as does 
the $c_2$ term in Eq. (1), and describes the spin-exchange interaction.
 
The free energy density of the ferromagnetic phase can be expanded in powers of 
$M^2$: $f_m = aM^2/2 + bM^4/4$, where $M$ is the 
order parameter of the ferromagnetic state and is proportional to the 
magnetization density. In the usual theory of phase transitions, $a$ changes 
sign at the ferromagnetic transition temperature and 
$b>0$\cite{note}. Here, we assume that $M$ is induced 
by the BEC itself, rather than by any other additional mechanisms, so $f_m$ 
should have a minimum at $M=0$. That is, the ferromagnetic state would 
disappear if it were not coupled to the Bose condensed state. Therefore, 
$f_m(M^2) \geq f_m(0)=0$ for all values of $M$, so both $a$ and 
$b$ are {\sl positive}. Near the critical temperature only the leading term 
need be considered, and $f_m$ takes the form, 
\begin{equation}
f_m = \frac {a}2 M^2 ~.
\end{equation}
Physically, $f_m$ should contain the energy of the magnetic field inside a 
magnetized medium\cite{book}. We suppose that the condensate only couples 
linearly to the ferromagnetic order, 
\begin{equation}
f_c = -g M \sum^F_{\sigma=-F} \sigma |\Psi_\sigma|^2 ~,
\end{equation}
where $g$ is the coupling constant. This is the simplest coupling term that 
satisfies the physical situation, as indicated later. Our discussions are 
confined to the thermodynamic limit and to one domain. Equations (4) and (5) 
comprise the basic ansatz of this work. 
We assume $a$ and $g$ are {\sl independent} of the temperature.

The total free energy for the Bose condensate can be written as 
$F_T=\int f_t dV$. Taking the variation with respect to 
$\Psi^*_\sigma$, $\Psi_\sigma$ and $M$, we derive
\begin{subequations}
\begin{eqnarray}
-\frac {\hbar^2}{2m} \nabla^2 \Psi_\sigma - (\alpha + gM\sigma)\Psi_\sigma 
   + \beta |{\bf \Psi}|^2 \Psi_\sigma  \nonumber \\ 
   + \beta_s \Psi^\dag_{\sigma'}{\bf F}_{\sigma\gamma} \cdot
     {\bf F}_{\sigma'\gamma'} \Psi_{\gamma'} \Psi_{\gamma} = 0~, \\
a M - g \sum^F_{\sigma=-F}\sigma |\Psi_\sigma |^2 = 0 ~.
\end{eqnarray}
\end{subequations}
Eq. (6a) consists of a set of equations with respect to the spin index 
$\sigma$. Eq. (6b) ensures that the magnetization is induced by the condensate 
itself.

Possible phase transitions of the spinor Bose gas can be investigated by 
solving Eq. (6). For a spatially uniform system, the first term in Eq. (6a) can 
be dropped. We take the bosons to have spin $F=1$. It has been suggested that 
the condensate should be ferromagnetic even {\sl without the spin-exchange 
interaction}\cite{eisenberg,yamada,Simkin,caramico}. To treat this case, we 
first set $\beta_s=0$. Two nontrivial solutions deserve further 
discussion\cite{note2}.

(I) {\sl Normal Bose-Einstein condensates}. This solution corresponds to BEC 
without spontaneous magnetization. 
\begin{equation}
|{\bf \Psi}|^2 = \frac {\alpha}{\beta} ~,~~ M=0 ~, 
\end{equation}
with the free energy density 
\begin{equation}
f^{I}_t = -\frac 12 \frac {\alpha^2}{\beta} ~.
\end{equation}
In this case, we have $|\Psi_1|^2 = |\Psi_{-1}|^2$ but cannot determine the 
proportion of $|\Psi_1|^2$ to $|\Psi_0|^2$. Thus, the system might be 
in ``polar", ``planar" or ``equal spin" states\cite{vollhardt}. 

(II) {\sl Bose-Einstein ferromagnets}. This solution describes BEC with a 
spontaneous magnetization:
\begin{eqnarray}
|\Psi_1|^2 = \frac {\alpha}{\beta-\frac {g^2}{a}} ~,~~
|\Psi_0|^2 = |\Psi_{-1}|^2 = 0 ~, \nonumber \\
M = \frac g{a} \frac {\alpha}{\beta-\frac {g^2}{a}} 
  = \frac g{a} |\Psi_1|^2 ~.
\end{eqnarray}
In this case the Bose condensate is fully polarized and the free energy 
density is
\begin{equation}
f^{II}_t = -\frac 12 \frac {\alpha^2}{\beta-\frac {g^2}{a}} ~.
\end{equation}

Comparing Eqs. (10) and (8), it can be seen that the BEF solution is indeed 
energetically favored in spite of a positive magnetization energy $f_m$. The 
magnetization results in a net energy decrease by $f^I_t-f^{II}_t 
\approx {g^2\alpha^2}/{(2a\beta^2)}$\cite{note3}, suggesting that $M \ne 0$  
occurs {\sl spontaneously}. This magnetization depends on the onset 
of BEC, with its magnitude proportional to the condensate density, as predicted 
in previous theories\cite{yamada,Simkin,caramico}.

As usual\cite{ginzburg}, $\alpha$ can be expanded in powers of $T_c-T$ near 
the Bose condensation temperature $T_c$: 
$\alpha(T) \cong \alpha^{\prime} (T_c-T)$; 
$\beta$ and $\alpha^{\prime}$ are positive constants. 
According to Eq. (9),
\begin{equation}
M =\frac g a |{\bf \Psi}|^2 = \frac g a |\Psi_1|^2 \propto T_c-T ~.
\end{equation}
The results are consistent with those obtained from a microscopic interacting 
boson model\cite{caramico}, suggesting that the phenomenological model 
describes the behavior of the spinor condensates quite well. 

The realization of BEC in optically trapped alkali-metal atoms makes the study 
on BEF not only theoretically interesting, but also experimentally attainable. 
We now turn our attention to such a specific BEC system. In this case, the 
$\beta_{s}$ term should be turned on. Using the notation 
$(\Psi_1, \Psi_0, \Psi_{-1})=
(\phi_1 e^{i \theta_1}, \phi_0, \phi_{-1} e^{i \theta_{-1}})$, the $\beta_{s}$ 
term becomes ${\beta_{s}} \phi^2_0 [ \phi^2_1 + \phi^2_{-1} +2 \phi_1\phi_{-1} 
\cos(\theta_1+\theta_{-1}) ] + 
\beta_{s} (\phi^2_1 - \phi^2_{-1})^2/2$\cite{isoshima}. 
Without considering BEF, only this term determines the relative phases of 
the three component condensate wave functions: 
$\theta_1+\theta_{-1}=\pi$ for $\beta_{s}>0$ and $\theta_1+\theta_{-1}=0$ for 
$\beta_{s}<0$. It therefore determines the ground state of the condensate: 
it is ferromagnetic for $\beta_{s}<0$ and ``polar" for $\beta_{s}>0$\cite{ho}. 
However, this phase diagram is modified by BEF. 
After some algebra, the free energy density reduces to 
$f^P_t = -{\alpha^2}/{(2\beta)}$ for ``polar" states and to 
$f^F_t = -{\alpha^2}/{[2(\beta+\beta_{s}-{g^2}/{a})]}$ 
for ferromagnetic states. Consequently, the phase diagram illustrated in Fig. 1 
results, with a phase boundary at a finite value of $\beta_{s}$, given by
\begin{equation}
(\beta_{s})_c = \frac {g^2}{a}~.
\end{equation}
Equation (12) implies that there is a critical $\beta_{s}$ value leading to the 
``polar" state.

Strictly speaking, the Landau expansion holds mathematically only near $T_c$. 
In order to get the {\sl ground state} phase diagram, we need decrease the 
temperature to zero, when the free energy density can be directly derived from 
the Hamiltonian via the Gross-Pitaevskii approximation. Because of 
the existence of BEF, we argue that {\sl the energy due to the magnetization}, 
as expressed in Eqs. (4) and (5), should be included in the $T=0$ free energy 
density functional. Then the Gross-Pitaevskii equation for the $F=1$ bosons 
described by Eq. (1) reads:
\begin{subequations}
\begin{eqnarray}
-\frac {\hbar^2}{2m} \nabla^2 \Psi_\sigma - (\mu + g^{*} 
   M\sigma)\Psi_\sigma + c_0 |{\bf \Psi}|^2 \Psi_\sigma  \nonumber \\ 
   + c_2 \Psi^\dag_{\sigma'}{\bf F}_{\sigma\gamma} \cdot
     {\bf F}_{\sigma'\gamma'} \Psi_{\gamma'} \Psi_{\gamma} = 0~, \\
a^{*} M - g^{*} \sum^F_{\sigma=-F}\sigma |\Psi_\sigma |^2 = 0 ~.
\end{eqnarray}
\end{subequations}
Here $\mu$ is the chemical potential. Equations (13) have similar forms as the 
phenomenological equations (6), but their parameters have {\sl different} 
meanings\cite{pitaevskii}, so here $a$ and $g$ are marked with a star. 
From Eq. (13), a phase diagram similar to 
Fig. 1 can be produced and the phase boundary becomes 
$(c_2)_c = {(g^{*})^2}/{a^{*}} ~.$ The Gross-Pitaevskii approximation is 
quite accurate in describing the ground state of Bose gases. 

\begin{figure}
\center{\epsfxsize=70mm \epsfysize=40mm \epsfbox{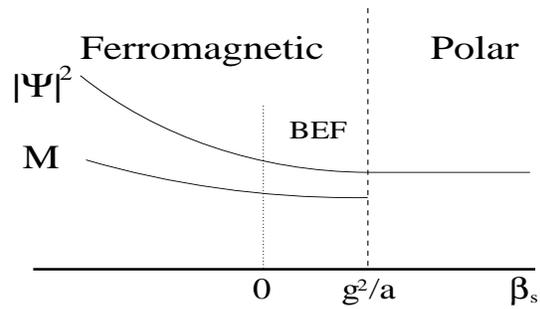}}
\caption{\label{fig:epsart}
Phase diagram of spin-$1$ Bose condensates. 
The schematic $\beta_{s}$-dependence of the two order 
parameters, $|{\bf \Psi}|^2$ and $M$, is outlined. Both decrease with 
increasing $\beta_{s}$ in the ferromagnetic state. At the phase boundary, 
$|{\bf \Psi}|^2$ is continuous but $M$ drops suddenly to zero, signaling a 
first-order phase transition. The intermediate region 
denotes the ferromagnetic state attributed purely to BEF. }
\end{figure}

We now discuss the origin of BEF. The possibility that BEF arises from an 
explicit spin-exchange interaction, like the $c_2$ term in Eq. (1), has been 
ruled out\cite{eisenberg,yamada,Simkin,caramico}. 
Our phenomenological model shows clues that BEF may be attributed to intrinsic 
magnetic moments of particles. Since spinor bosons carry magnetic moments 
${\bf m}$, an effective magnetic field ${\bf B}=\mu_0 {\bf M}$ is created 
inside a polarized body, owing to the superposition of the intrinsic field of 
all the magnetic moments. $\mu_0$ is the permeability of vacuum and 
${\bf M}={\bf m} |\Psi_1|^2$ is the magnetization. 
Naturally, $f_m={B^2}/{(2\mu_0)}={\mu_0} M^2/2$ can be regarded as 
the energy of the effective field. On the other hand, 
the magnetic moment of the condensed bosons does respond to the magnetic 
field regardless if the field is external or internal, so the spin direction 
should follow the internal field ${\bf B}$. Eq. (5) just is the general 
form of the coupling between the condensate density and the internal magnetic 
field: $f_c=-{\mu_0} m M |\Psi_1|^2=-{\bf M\cdot B}=-\mu_0 M^2$. Within this 
interpretation, $f_m+f_c = -{\bf \mu_0} M^2/2$, which is just the self-energy 
of a magnetized body with magnetization ${\bf M}$\cite{book}. Contrarily, in 
the ``polar" or ``equal spin" states the magnetic moments of the particles are 
fully compensated: $M=0$. Then at the 
mean-field level, an internal reference particle cannot sense the moments of 
other particles, and the magnetic self-energy is zero. Obviously the 
ferromagnetic state is energetically favorable at $T=0$. At finite temperature, 
the ferromagnetic state can be destroyed due to an entropy increase. However, 
once Bose-Einstein condensation takes place, the temperature inside a 
condensate is absolutely zero and the ferromagnetic state remains stable. This 
accounts for the dependence of BEF on the onset of BEC. 
We note that such a interpretation coincides with the prediction of Yamada and 
Caramico D'Auria {\sl et al.}, because in their calculations of magnetization 
at a given field, the magnetic moment of spinor bosons was included. 

Simkin and Cohen attributed BEF to the Bose-Einstein statistics\cite{Simkin}. 
However, we notice that although the Bose-Einstein statistics does not exclude 
a ferromagnetic state in spinor bosons, it cannot lead to a spontaneous 
polarization. According to EL, the ``fully" polarized state is among the 
degenerate ground states but the degeneracy cannot be lifted spontaneously if 
we do not consider the magnetic moment of particles.
Rojo\cite{rojo} studied a two-component bosonic system which can be mapped 
into a ``pseudo-spinor" Bose gas with $F=1/2$ (without magnetic moments) and 
showed that the ground state is not polarized when the interaction is 
component-independent.

Within the interpretation that BEF is due to the magnetic moment, the 
{\sl magnetic dipole interaction} between particles is implicitly involved. The 
dipole interaction is caused by the intrinsic field of magnetic moments 
which forms the internal magnetic field when magnetic moments are arranged 
parallel. The influence of magnetic dipole interactions on the properties of 
BEC has attracted considerable interest recently\cite{note4}. Although very 
weak, they can modify the ground state and the collective excitations of 
trapped condensates significantly\cite{pu,note4}. Our approach gives a new 
perspective for studying the effect of the dipole interactions on the boson 
gas. It is worth noting that dipole interactions can hardly result in a 
ferromagnetic state in a real Fermi gas, because the interaction is too weak 
to approach the Stoner threshold.

The intermediate area in Fig. 1 denotes the ``pure" Bose-Einstein 
ferromagnetic state. This provides a direct way of detecting BEF 
experimentally. Once we have one experimental example pertaining to this 
regime, i.e., $c_2>0$ and 
the ground state is still ferromagnetic, the existence of 
BEF could be confirmed. The parameter $c_2\propto (a_2-a_0)/3$ can be 
estimated from experiments\cite{ho}, where $a_2$ and $a_0$ are s-wave 
scattering lengths corresponding to the total spin two and zero channel, 
respectively. According to the above arguments, 
\begin{equation}
(c_2)_c = \mu_0 {\bf m}^2,
\end{equation} 
It is a rather small value. This makes the direct examination of BEF a 
difficult task.

Up to now, all of our discussions are devoted to chargeless bosons. One can 
expect that charged bosons exhibit more complex properties. For example, 
there would be a competition between the Bose-Einstein ferromagnetism and the 
Meissner effect, even though there is no external magnetic field applied. The 
latter tends to reduce or even to kill the magnetization. So, the nature of the 
ground state for charged spinor bosons is in our view still an open question. 
Relevant topics have been studied in triplet superconductors\cite{Scharnberg}. 

In conclusion, we have proposed a phenomenological model consistent with the 
prediction that magnetic bosons undergo a spontaneous 
magnetization together with Bose-Einstein condensation. We remark that this 
phenomenon deserves special attention due to two reasons: 
(i) It is an intrinsic property of spinor bosons and may be attributed to the 
magnetic moment of particles. (ii) The spontaneous magnetization is parasitic 
to the Bose-Einstein condensation, reflecting that the entropy of the Bose 
condensate is zero.

QG is grateful to Dimo I. Uzunov for his guidance and valuable discussions and 
to Richard A. Klemm and J. Brand for their helpful suggestions and careful 
reading of the manuscript. Useful discussions with Y.-J. Wang and V.P. Mineev 
are also acknowledged.


\begin{references}

\bibitem{cornell} M.H. Anderson {\sl et al}., Science {\bf 269}, 198 (1995);
K.B. Davis {\sl et al}., Phys. Rev. Lett. {\bf 75}, 3969 (1995);
C.C. Bradley {\sl et al}., {\sl ibid}. {\bf 75}, 1687 (1995).

\bibitem{leggett} See for reviews, A.J. Leggett, Rev. Mod. Phys. {\bf 73}, 
307 (2001) and F. Dalfovo {\sl et al}., {\sl ibid}. {\bf 71}, 463 (1999)

\bibitem{ketterle} D.M. Stamper-Kurn {\sl et al}., 
Phys. Rev. Lett. {\bf 80}, 2027 (1998).

\bibitem{kett2} J. Stenger {\sl et al}., Nature (London) {\bf 396}, 345 (1998).

\bibitem{ho}  T.L. Ho, Phys. Rev. Lett. {\bf 81}, 742 (1998).

\bibitem{ohmi} T. Ohmi and K. Machida, J. Phys. Soc. Jpn. {\bf 67}, 1822 (1998).

\bibitem{pu} H. Pu, W. Zhang, and P. Meystre, Phys. Rev. Lett. {\bf 87}, 
140405 (2001); W. Zhang {\sl et al}., {\sl ibid}. {\bf 88}, 060401 (2002); 
K. Gross {\sl et al}., Phys. Rev. A {\bf 66}, 033603 (2002).

\bibitem{eisenberg} E. Eisenberg and E.H. Lieb, Phys. Rev. Lett. {\bf 89}, 
220403 (2002).

\bibitem{yamada} K. Yamada, Prog. Theor. Phys. {\bf 67}, 443 (1982).

\bibitem{Simkin} M.V. Simkin and E.G.D. Cohen, Phys. Rev. A {\bf 59}, 
1528 (1999).

\bibitem{caramico} A. Caramico D'Auria, L. De Cesare, and I. Rabuffo, 
Physica A {\bf 225}, 363 (1996).

\bibitem{suto} A. S\"{u}t\H{o}, J. Phys. A {\bf 26}, 4689 (1993).

\bibitem{ginzburg} V.L. Ginzburg and L.P. Pitaevskii, 
Sov. Phys. JETP {\bf 7}, 858 (1958).

\bibitem{note} The Landau free energy in this form describes a spontaneous 
magnetization caused by other mechanisms. One can have $f_m$ in this form 
to study the interplay of the ferromagnetic and superfluid states. But that is 
not the purpose of this work.

\bibitem{book} D.J. Craik, {\sl Magnetism: principles and applications} 
(John Wiley \& Sons, England, 1995).

\bibitem{note2} There are other solutions with higher free energies, including 
a trivial solution $|\Psi|^2=0$, $M=0$ and a decoupled solution 
$|\Psi_1|^2=|\Psi_{-1}|^2=0$, $|\Psi_0|^2=\alpha/\beta$ with $M$ remaining 
arbitrary.

\bibitem{vollhardt} D. Vollhardt and P. W\"olfle, {\sl The superfluid phases of 
Helium 3} (Taylor \& Francis, London, 1990).

\bibitem{note3} Herein we suppose $\beta \gg {g^2}/{a}$. The free energy 
density corresponding to solution (II) can be written as $f^{II}_t = 
- \alpha |\Psi_1|^2 + ( \beta - {g^2}/{a} )|\Psi_1|^4/2$. If 
$\beta<{g^2}/{a}$, higher-order expansions, e.g. $|{\bf \Psi}|^6$, were to be 
considered, the transition might be first order.  

\bibitem{isoshima} T. Isoshima, K. Machida, and T. Ohmi, 
Phys. Rev. A {\bf 60}, 4857 (1999).

\bibitem{pitaevskii}  L.P. Pitaevskii, Sov. Phys. JETP {\bf 13}, 451 (1961).

\bibitem{rojo} A.G. Rojo, Phys. Rev. A {\bf 64}, 033608 (2001).

\bibitem{note4} See for example, K. Goral, K. Rzazewski, and T. Pfau, 
Phys. Rev. A {\bf 61}, 051601 (2000); J.-P. Martikainen, M. Mackie, and 
K.-A. Suominen, {\sl ibid}. {\bf 64}, 037601 (2001); K. Goral, L. Santos, and 
M. Lewenstein, Phys. Rev. Lett. {\bf 88}, 170406 (2002); 
S. Giovanazzi, A. G\"orlitz, and T. Pfau, {\sl ibid}. {\bf 89}, 130401 (2002).

\bibitem{Scharnberg} K. Scharnberg and R.A. Klemm, Phys. Rev. B {\bf 22}, 
5233 (1980).

\end{references}
\end{document}